\documentclass{aa}  

\usepackage{natbib}
\usepackage{graphicx}
\usepackage{babel}
\usepackage{caption}
\usepackage{subfig}
\usepackage[varg]{txfonts}
\usepackage{rotating}
\usepackage{dcolumn}
\usepackage{fmtcount}
\usepackage{longtable, booktabs}
\usepackage{array,multirow}
\usepackage{color}
\usepackage{aas_macros}
\usepackage{multirow}
\bibpunct{(}{)}{;}{a}{}{,}
\usepackage{hyperref}
\usepackage{xcolor}
\usepackage{ulem}
\definecolor{dark-red}{rgb}{0.5,0.15,0.15}
\definecolor{dark-blue}{rgb}{0.15,0.15,0.5}
\definecolor{medium-blue}{rgb}{0,0,0.5}
\definecolor{medium-red}{rgb}{1,0,0}
\hypersetup{
    colorlinks={true}, linkcolor={medium-red},
    citecolor={dark-blue}, urlcolor={medium-blue}
}

\newcommand{\vrad}{\ensuremath{v_{\mathrm{rad}}}\xspace}
\newcommand{\teff}{\ensuremath{T_{\mathrm{eff}}}\xspace}
\newcommand{\kms}{\ensuremath{\rm{km}\,s^{-1}}\xspace}
\newcommand{\logg}{\rm{log g}\xspace}
\newcommand{\feh}{\rm{[Fe/H]}\xspace}
\newcommand{\afe}{\rm{[$\alpha$/Fe]}\xspace}

\defcitealias{husser16}{H16} 	  	 
\newcommand{\TOH}{\citetalias{husser16}\xspace}

\usepackage{xcolor,ulem}
\newcommand\hl{\bgroup\markoverwith{\textcolor{yellow}{\rule[-.5ex]{2pt}{2.5ex}}}\ULon} 
\definecolor{dark-green}{rgb}{0.10,0.4,0.1}

\begin{document} 

   \title{NGC\,6397: The metallicity trend along the isochrone revisited}

   \author{
     Rashi Jain \inst{1}
     \and
     Philippe Prugniel \inst{2}
     \and
     Lucimara Martins \inst{3,2}
     \and
     Ariane Lan\c con \inst{1}
   }
   \institute{
     Universit\'{e} de Strasbourg, CNRS, Observatoire astronomique de Strasbourg, UMR7550, F-67000, Strasbourg, France
     \and
     Centre de Recherche Astrophysique de Lyon UMR5574, Univ Lyon, Univ Lyon1, ENS de Lyon, CNRS, F-69230 Saint-Genis-Laval, France
     \and
     NAT- Universidade Cidade de S\~ao Paulo/Universidade Cruzeiro do Sul, Rua Galvão Bueno, 868, 01506-000 Sao Paulo, Brazil
   }
   \date{Received xxxx; accepted xxxx}


  \abstract
{  Recent work has used spectra of $\sim$5000 stars in NGC\,6397 that were extracted from a MUSE mosaic to determine the atmospheric parameters for these stars by fitting the spectra against the G\"ottingen Spectral Library. A significant change in metallicity between the turn off and the red giant branch was found  and was discussed as a possible manifestation of predicted effects of atomic diffusion. However, the small amplitude of the effect and inconsistency with earlier measurements call for more attention before this result is interpreted. Systematic effects due to the interpolation or to the synthetic spectra cannot be ruled out at this level of precision.
}
      {We reanalyze the data with : the ELODIE and MILES reference libraries in order to assess the robustness of the result. These empirical libraries have a finer metallicity coverage down to approximately the  cluster metalicity turn-off.
   }
   {
     We used the ULySS full-spectrum fitting package, together with the library interpolators to remeasure the three atmospheric parameters effective temperature, surface gravity, and \feh metallicity.
   }
   { We find a very low \feh dispersion along the isochrone (0.07 dex), consistent with our error estimate (0.05 dex). 
However, the \feh trend is not reproducible.
This shows that the data have the potential to reveal patterns of the magnitude of the expected physical effects, but the analysis methods need to be refined to cull systematic effects that currently dominate the patterns.
}
   {}


   \maketitle

\section{Introduction}

The introduction of the wide-field integral field spectrograph MUSE \citep{bacon2010,bacon2014} mounted at UT4 (Yepun) of the ESO Very Large Telescope (VLT) recently ushered in a new area for the exploration of Galactic globular clusters. \citet[hereafter H16]{husser16} carried out and analyzed observations of NGC\,6397. With a field of one arcminute and a spatial resolution limited by the seeing, MUSE allows mapping a full globular cluster
in a reasonable number of pointings.
The instrument allows obtaining many more spectra in one shot than a multi-object spectrograph, and the continuous mapping of the area makes it possible to use deblending methods similar to those that are commonly applied for crowded-field photometry \citep[e.g.,  DAOPHOT,][]{stetson1987}.
Although these spectra do not have the high spectral resolution required for abundance analysis \citep{alves2012,ernandes2018}, they are suitable for determining the atmospheric parameters, effective temperature (\teff), surface gravity (\logg), and metallicity (\feh) using methods that are applicable at low resolution. Full-spectrum fitting is one of these methods, and its reliability and efficiency has been demonstrated in several instances. For example, \citet{koleva12} have shown that the parameters can be retrieved from spectra with a resolution $R = \lambda/\Delta\lambda \approx 1000$, and the authors found that the precision was not degraded compared to high-resolution spectra or the appearance of significant biases. Full spectrum fitting offers the advantage of optimally using the entire signal, which is particularly valuable in the present case, where the observations cover a range of apparent magnitudes and hence of signal-to-noise ratios (S/N).

\TOH analyzed 18\,932 spectra of 12\,307 stars extracted from a mosaic of 23 MUSE pointings with full-spectrum fitting. The observations were compared to a grid of models \citep[the {\it G\"ottingen Spectral Library}, GSL]{husser13} that were computed with the PHOENIX synthesis code \citep{allar95} which was interpolated to minimize the residuals. One of the remarkable results of this work is the finding of a metallicity trend along the isochrone, with an amplitude of 0.2 to 0.3 dex. The metallicity is minimum at the turn-off (TO) and rises both along the sub-giant branch (sGB) and down the main sequence (MS).
This is reminiscent of the expectations from atomic diffusion, which reduces the surface abundances of metals in regimes where mixing processes such as turbulence are relatively inefficient \citep{richard2002}. 

The mean observed effect in \TOH is slightly stronger than found in the
detailed abundance analysis of individual cluster stars
of \cite{nordlander2012}, which was $\sim 0.1$\,dex on average, but
was based on a small number of stars in each of the evolutionary phases
of interest.
The authors of this study highlight the sensitivity of the metallicity trend to model aspects such as the efficiency of the turbulent mixing, temperature scale, nonlocal thermal equilibrium (NLTE) and 3D effects.
The model grid used in \TOH was computed with 1D models.

\begin{figure*}
\begin{center}
	\includegraphics[width=1\textwidth, trim=0.3in 0.in 0.in 0.in]{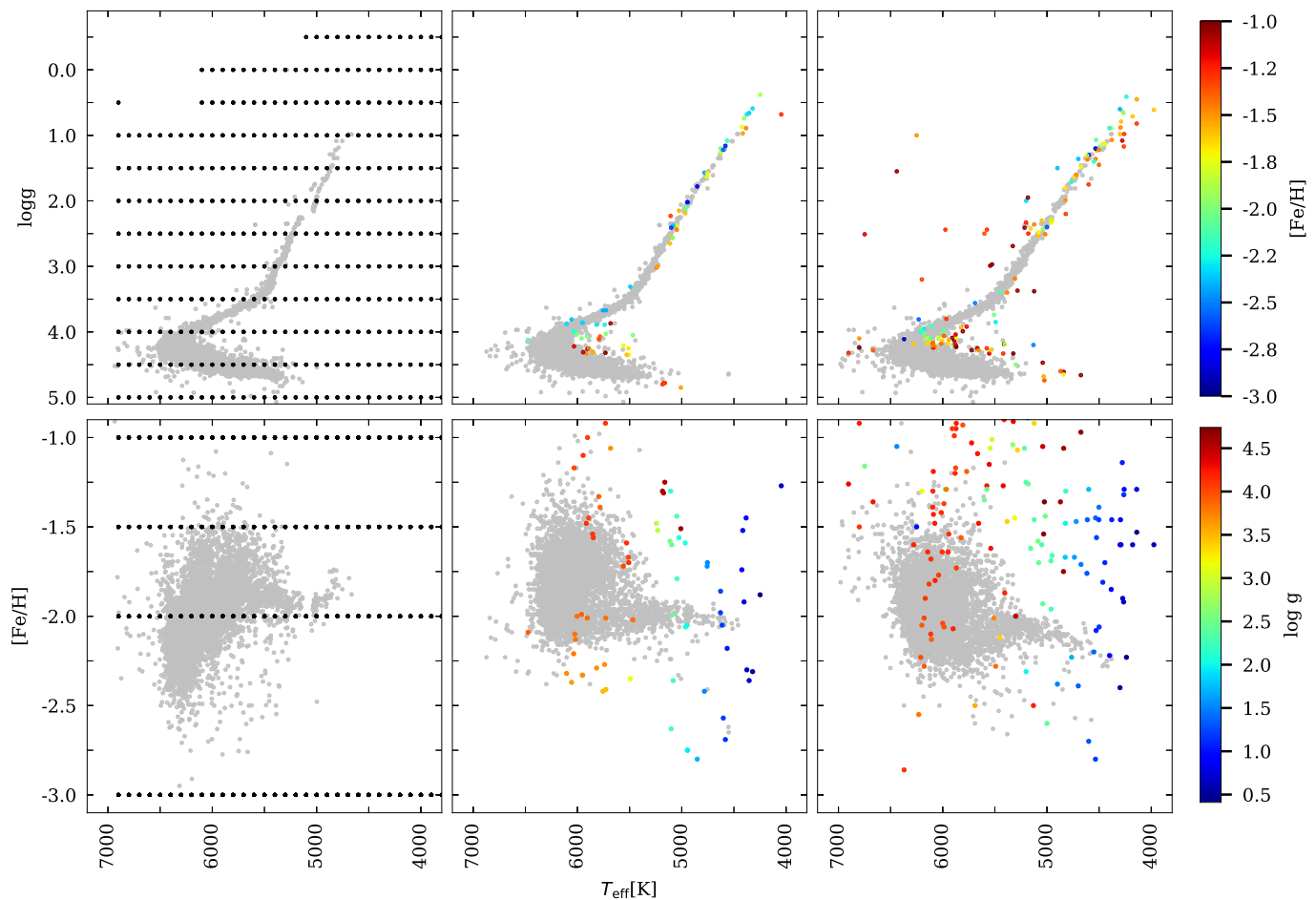}
	\caption{Distribution of the library spectra in the region of the parameter space occupied by the cluster stars. The gray dots represent the \TOH measurements of the cluster members. The top graphics are the projections in the \logg vs. \teff plane and the bottom graphics show the \feh vs. \teff projections.
          The overplotted colored points are the location of the reference spectra in the three libraries used in this paper. The left panel shows the {\it G\"ottingen Spectral Library} used by \TOH, the central panel shows the ELODIE library, and the right panel the MILES library. These last two libraries were used in tour reanalysis. Only the library stars corresponding to the region mapped by the our sample are shown: $\teff < 7000$ K, and $-3.04 < \feh < -0.88$ dex (77 stars for ELODIE and 137 for MILES, including 40 and 46 stars, respectively with $\feh < -1.7$~dex).
The color scales for the ELODIE and MILES libraries are indicated on the right. 
          }
	\label{fig:libstars}
\end{center}
\end{figure*}

In addition, the magnitude of the detected
metallicity  variations are only a fraction of the mesh of the grid of
available reference synthetic spectra.
The GSL was computed at regular spacing in the \teff, \logg, \feh, and \afe parameter space, and the steps in metallicity are 1.0 dex for $\feh < -2.0$ and 0.5 dex for $\feh > -2.0$. Figure~\ref{fig:libstars} (left panel) shows the distribution of the GSL spectra in the \logg versus \teff and \feh versus \teff diagrams, plotted over the distribution of the individual stars of the cluster. The mean metallicity of the cluster is close to the $\feh = -2.0$ dex slice of the GSL, and the whole metallicity range is well within the two surrounding slices, at $-1.5$ and $-3.0$ dex.
We connot rule out that the measured trend is affected by the spline interpolation in the grid.

The potential significance of the result for our understanding of
mixing processes in stars, may affect quantities of general interest
in the study of star clusters, such as stellar lifetimes or the location
of the TO. We therefore here revisit the result using a different approach.
Studies of the integrated light of globular clusters have shown
that the residuals of full spectrum fits are significantly lower when
the artificial cluster spectra are based on empirical spectral libraries
than when they are based on theoretical libraries \citep{martins2019}.

We use two of these empirical spectral libraries to
reanalyze the stellar spectra of the MUSE observations of NGC\,6397.
By using other reference spectra than \TOH, we introduce a different set
of biases.  Differences with the results of \TOH will
provide an assessment of systematic uncertainties, and the basis
for future improvements.

Section \ref{sec:data} presents the data and their analysis, Sect. \ref{sec:analysis} our new analysis, Sect. \ref{sec:discuss} discusses the results, and Sect. \ref{sec:conclusions} draws the conclusions.


\section{Data and earlier analysis}\label{sec:data}

The observations of a 5 x 5 mosaic\footnote{Two of the planned fields, located at the periphery, were not observed; 23 fields are available.} reaching out to a distance of $\sim$ 3.5 arcmin from the cluster center were made during the MUSE commissioning, between 2014 July 26 to August 3 (ESO program identifier {\tt 60.A-9100(C)}).
They consisted of 127 pointings of exposure time of not more than 60 s to avoid saturation of the bright giant stars, for a total integration time of 95 min. Each pointing covers a field of 1 x 1 arcmin with a spatial step of 0.2 arcmin, and the seeing was in the range 0.6'' to 1''.  

The data were reduced by the MUSE consortium using the official pipeline, and the extraction of the individual stellar spectra was described in \TOH, with a deblending approach presented in \citet[implemented in the PampelMuse software, publicly available]{kamann2013}. This method relies on a photometric and astrometric input catalog, which in this particular case was the \citet{anderson2008} table derived from images obtained with the Advanced Cmaera for Surveys (ACS) on board the Hubble Space Telescope (HST). An initial point spread function (PSF) model, set to be the seeing, is used to create a mock image that is cross-correlated with each wavelength layer of the MUSE cube to find the coordinates transformations. The most isolated and bright spectra are used to iteratively refine the PSF by subtracting all the other spectra. The final wavelength-dependent PSF is derived by smoothing the PSF obtained for each layer, and it is used to extract the spectra. In total, 18\,932 spectra of 12\,307 stars were extracted. The stars in the overlapping regions between the different fields were observed multiple times.
The line spread function (LSF) varies with the wavelength from a full-width at half-maximum (FWHM) of about 2.82 \AA{} at $\lambda = 4750 $\AA{} (equivalent to an instrumental velocity dispersion, $\sigma_{ins} = 80$ \kms) to about 2.54 \AA{} at $\lambda = 7000 $\AA{} ($\sigma_{ins} = 46$ \kms), and stays approximately constant further in the red. The actual LSF also varies across the 24 spectrographs forming MUSE by roughly 0.1 \AA.

Out of the 18\,932 extracted spectra, 14\,271, with an SNR greater than 5 are distributed on the MUSE website\footnote{http://muse-vlt.eu/science/globular-cluster-ngc-6397} (the raw data are available in the ESO archive\footnote{http://archive.eso.org/}).

To determine the atmospheric parameters, \TOH proceeded in three steps. First they obtained \teff and \logg by fitting the HST photometry to an isochrone. They then used these parameters to generate interpolated models that they cross-correlated with the observations to derive the radial velocity (\vrad).
In the third step, they performed an optimization to produce the final \teff, \feh, \vrad, line broadening, and telluric absorption spectrum.
They used their own full-spectrum fitting procedure, using GSL interpolated by cubic spline as reference. The interpolation cannot be a simple linear interpolation because the optimization method requires its derivative to be continuous.
The value of \logg was not optimized; the adopted value is the photometric one. The authors justify this choice as a precautionary approach to prevent a possible degeneracy between \logg and the broadening.

This analysis provided parameters for 5\,882 spectra of 4\,132 stars with an S/N $> 20$, above which the formal error on \teff is lower than 100 K, and on \feh lower than 0.16 dex. The tables with results were kindly provided to us by Tim-Oliver Husser. For 367 of these spectra \logg was not determined, and consequently, neither \teff nor \feh were measured. 
To clean the sample of the nonmember stars, we selected the spectra within an ellipse centered on \vrad = 17.8 \kms and \feh = -1.96 dex 
(mean parameters of the cluster), and semiaxes of 29 \kms and 1.08 dex (corresponding approximately to the selection area measured in Figure 7 of \TOH; the actual values are not reported in the text of this paper).
The sample is reduced to 5\,510 spectra of 4\,089 distinct cluster members. Furthermore, we are interested here in the TO and red giant branch (RGB), and because the method of determining the stellar parameters by full-spectrum fitting has never been validated for the hot stars, we followed \TOH and limited the sample to $\teff < 7000$ K. This criterion excludes the hot horizontal branch stars. Finally, the sample contains  5\,400 spectra of 4\,053 distinct cluster members.

\section{Reanalysis}\label{sec:analysis}

\subsection{Method}

The aim of this paper is to repeat \TOH's analysis of the metallicity trend along the cluster sequence with other reference libraries in order to determine the reliability of this trend. Alternatives to GSL are the ELODIE and MILES empirical libraries, among others.
Synthetic and empirical libraries are seen as complementary; each has their advantages and drawbacks. On the one hand, synthetic spectra can be computed at any point in the parameter space and are therefore not limited to the region populated with stars that can be observed (i.e., essentially stars from the solar neighbourhood). They are also free of noise and observational signatures, and their spectral resolution can be as high as necessary. However, synthetic spectra still fail to accurately match observed spectra, because the physical models and lists of atomic and molecular transitions are limited \citep{martins2014}.
On the other hand, empirical libraries can accurately reproduce observed spectra, and this is why they are extensively used to model integrated spectra of stellar populations \citep{martins2019}.

ELODIE is an empirical library, initially presented in \citet{prugniel2001} and later upgraded to increase the coverage in the parameter space. We used the latest version (version 3.2), described in \citet{wu2011}, containing 1962 spectra at a constant FWHM resolution of 0.55 \AA{} over the wavelength range 3900 -- 6800 \AA{} (R $\approx$ 10\,000). The second library, MILES \citep{sanchez-blazquez08}, contains spectra of 985 stars at a resolution about 2.55 \AA{} (R $\approx$ 2000) in the wavelength range 3536-7410 \AA.
The distribution of the stars populating the region of the cluster in the parameter space are represented in the central and right panels of Fig.~\ref{fig:libstars} for ELODIE and MILES, respectively. Both libraries sample the region of the RGB reasonably well down to almost the TO.
Along the MS, most of the stars are on the high-metallicity side compared to the cluster, and only a few have the cluster metallicity.
It is also noticeable that the library MSs are shifted toward lower gravities by about 0.1~dex compared to the cluster star gravities determined by isochrone fitting. We found indications that this may be at least partly due to an underestimate of the gravities in the libraries.

The fitting procedure to be used for estimating stellar parameters requires interpolating through the libraries to compute a spectrum for any set of \teff, \logg, and \feh. For this purpose, we used polynomial interpolators computed using stellar parameters of the library stars compiled from the literature, generally obtained through detailed abundance studies using high-resolution spectroscopy.
The ELODIE interpolator is described in \citet{wu2011}, and 
for MILES, we used the interpolator described in \citet{sharma16}.
The quality of the interpolated spectra naturally depends on the distribution of the stars in the parameter space: the denser the distribution, the better the interpolation. The quality also depends on the accuracy of the input stellar parameters that were used when the interpolators were computed, and on the intrinsic accuracy of the interpolators. At the margin of the parameter space, the interpolators use synthetic spectra to extend the validity range, but with a lower quality.
The MS of the cluster is at the edge of the validity region of the interpolators. However, using the procedure described below, we successfully fit synthetic spectra from \citet{coelho2005} and from the GSL that were chosen to map this regime. This gave us some confidence in our ability to analyze MS spectra. However, similar tests with the cluster's spectra failed to restore the correct gravity. The solutions were found at a correct temperature, but located on the giant branch. The reasons for this difficulty are not completely clear, but are probably related to the drop in S/N from $\sim$60 near the TO to $\sim$20 on the lower MS, and to stronger contamination by other cluster stars for the fainter objects. We here prefer to restrict the spectral analysis to \logg<4.2 and high S/N. and are subject to a strong light contamination by the other cluster stars.
Although we here follow \TOH and adopt the isochrone gravities, we prefer to restrict our analysis to the RGB, $\logg < 4.2$. This sample contains 1\,587 spectra of 1\,063 stars.

The atmospheric parameters were determined with the ULySS package \citep{koleva09} by minimizing the squared residuals between the MUSE observation and an interpolated spectrum,

\begin{equation} \label{eqn:ulyss}
    S(\lambda) = P_n(\lambda) \times G(\vrad, \sigma_{\mathrm{rel}}) \otimes \mathrm{TGM}(\teff, \logg, \feh,\lambda),
\end{equation} 
\noindent where $P_n(\lambda)$ is a series of Legendre polynomial up to degree $n$, meant to absorb the instrumental spectral response and line-of-sight extinction; we used $n=20$.
$G(\vrad, \sigma_{\mathrm{rel}})$ is a Gaussian function centered at \vrad and with standard deviation  $\sigma_{\mathrm{rel}}$. The spectra were binned in logarithmic wavelength, so that the Doppler shift can be expressed by a convolution.
TGM is the spectral interpolator. 
The free parameters are \teff, \feh, \vrad, $\sigma_{\mathrm{rel}}$ and the coefficients of $P_n(\lambda)$. 
See \citet{arentsen2019} for other details on the procedure.

$\sigma_{\mathrm{rel}}$ was let free in order to account for the variation of instrumental broadening from spectrum to spectrum (as described 
in Sect.~\ref{sec:data}, the MUSE LSF varies between the individual spectrographs). The relative LSF between MUSE and the libraries changes significantly over the wavelength range, and we therefore ingested this variation pattern in the interpolated spectra, before the minimization. This was made with a stepwise convolution of the library (see the {\sc uly\_lsf\_convol} function in the ULySS package). The LSF injection improves the quality of the fits and slightly reduces the errors of the estimated parameters, but it does not affect the general trends.

In the case of the ELODIE library, the injected LSF was chosen so that the interpolated spectrum has a higher resolution than the MUSE spectrum. The observation was then fit by letting $\sigma_\mathrm{rel}$ free.
For the MILES library, the resolution after the LSF injection is very close to that of MUSE, but because of the spectrum-by-spectrum variations by 0.1 \AA{}, it is sometimes lower. In such cases, detected because the first fit fails to determine $\sigma_\mathrm{rel}$, we convolved the observation with a Gaussian with a dispersion of 30 \kms, and then let the fitting procedure determine $\sigma_\mathrm{rel}$.

A difference between the present procedure and that of \TOH is that the latter adjusted the telluric absorption together with the stellar parameters and broadening. Telluric absorption features are prominent in the red part of the spectra, which are outside the currently analyzed region, and we more simply masked (i. e., ignored) the two strongest features. The first feature is the water absorption near the NaD line that affects both the ELODIE and MILES, and the second is the B band, near 6887 \AA{} \citep{f1817,f1821}, which affects MILES.

No error spectra are associated with the extracted observations, but a  mean S/N is provided. We used it to  determine a noise level that we assumed to be identical throughout the spectrum, meaning that all wavelength bins have the same weight. This assumed error sets the scale of the fitting errors on the parameters that we tune in Sect.~\ref{sec:internal_errors} into internal errors, by matching the repeated observations of stars in the overlaps of the different pointings.

In order to keep the reanalysis as closely as possible to that of \TOH, we carried it out by fixing \logg to their photometric value (but see Sect.~\ref{sec:freeg}, where we also determine \logg spectroscopically).
As first guesses, we used for \teff, the photometric value, and for \feh we used -2.0 dex, which is an approximation of the cluster metallicity.

\subsection{Assessment of the internal errors} \label{sec:internal_errors}

The full catalog contains 5\,510 measurements for 4\,089 stars (including the stars from the MS to the RGB).
For the 1\,200 stars with two or more repeated observations, the spectra differ by the noise and by the propagation of systematics introduced by the spectrograph and data reduction. We assessed the precision by comparing the atmospheric parameters measured on each pair of observations as in \citet{arentsen2019}. We computed for each pair of observations $i$:
\begin{equation}
    \Delta \mathrm{P}_{i} = \frac{\mathrm{P}_{1,i} - \mathrm{P}_{2,i}}{\sqrt{\epsilon_{1,i}^2 + \epsilon_{2,i}^2}},
    \label{eqn:P}
\end{equation} 

\noindent where P can be \teff or \feh, $\epsilon$ is the formal ULySS error on the respective parameters, and $1$ and $2$ indicate the observation of each spectrum of the pair. This distribution is expected to be nearly Gaussian, and if the errors are properly scaled, their standard deviation is expected to be one. For both parameters we find $\Delta \mathrm{P} =$ 2.2 and 1.8 for ELODIE and MILES, respectively. Because we assumed a constant noise scaled to produce the average S/N determined by the data reduction and spectrum extraction pipeline, we indeed expected deviations from unity. We also assumed that all the pixels are independent, which cannot be the case because the spectra were rebinned into evenly sampled wavelength (this process necessarily introduces 
a correlation in the noise), and this leads to an underestimation of the errors.

We verified that these distributions were similar throughout the entire parameter space, and we found only a marginal indication that they may depend on S/N, in a way consistent with an additional source of noise independent of S/N. The effect is about 15 K and 0.01 dex to be quadratically added on the errors on \teff and \feh,respectively. This is very small.

The fact that the distributions are similar for the two parameters is a satisfactory sanity check of our analysis, and we did not investigate the deviations from unity further because they are not relevant for our present discussion. We rescaled the errors by the factors described above when we computed the errors in the final table (see Sect.~~\ref{sec:results}).

\subsection{Assessment of the external errors (accuracy)}\label{sec:external}

For an estimate of the external errors, we compared our measurements with previous works. 
First of all, the comparison of 1587 measurements with \TOH indicates $\teff({\rm\scriptstyle H16}) - \teff({\rm\scriptstyle ELODIE/MILES}) = 92$~K with a dispersion of $40$~K.
For the metallicity, the bias is different for ELODIE or MILES:  $\feh({\rm\scriptstyle H16}) - \feh({\rm\scriptstyle ELODIE}) = -0.07$~dex, and  $\feh({\rm\scriptstyle H16}) - \feh({\rm\scriptstyle MILES}) = 0.03$, with a dispersion of 0.15 dex.
Because the three series used the same reduced spectra, they can be similarly affected by systematics due to the data. These figures were therefore be regarded as lower limits to the errors, and comparisons with independent measurements will give a more complete assessment of the external errors.

Five previous studies determined the parameters of a significant number of stars in this cluster.
\citet{carretta2009a} measured 13 stars at high spectral resolution, with the UVES spectrograph attached to the VLT, and \citet{carretta2009b} measured 144 stars at intermediate resolution, with GIRAFFE on the same telescope, all giants with $\teff < 5400$~K. \citet{lovisi2012} measured 146 stars, but those in common with the MUSE sample are mostly blue stragglers, which are not discussed here. Finally, \citet{korn2007}, and \citet{lind2008} observed stars with the UVES (18 stars) and GIRAFFE (116 stars) spectrographs, respectively, but
in more peripheral fields that lie outside the MUSE pointings.

The comparisons with the two first datasets are very similar, so that we can merge them and report only the total comparison.  We find $\teff({\rm\scriptstyle lit}) - \teff({\rm\scriptstyle ELODIE}) = 15$~K with a dispersion of $30$~K, and  $\teff({\rm\scriptstyle lit}) - \teff({\rm\scriptstyle MILES}) = 48$~K with a dispersion of $25$~K. For the metallicity, we find no differences with ELODIE, and  $\feh({\rm\scriptstyle lit}) - \feh({\rm\scriptstyle MILES}) = 0.07$~dex, in both cases with a dispersion of $0.05$~dex.

The comparison between the \TOH measurements and the literature displays the following differences:  $\teff({\rm\scriptstyle lit}) - \teff({\rm\scriptstyle H16}) = -100$~K and $\feh({\rm\scriptstyle lit}) - \feh({\rm\scriptstyle H16}) = -0.08$~dex, with dispersions of $48$~K and $0.06$~dex on the two parameters. In all cases the \logg values are very consistent with the literature, which is expected because they were determined photometrically in all cases with very similar methods.

The picture that emerges from these comparisons is that (i) \teff measured by \TOH is on average hotter than the literature and our own measurements. (ii) \feh is reasonably consistent between the different series; our measurements are marginally more consistent with the literature than those of \TOH. (iii) The consistency is similar for our two analyses, with a marginally better performance of ELODIE.

For the observations compared with the literature, the mean estimated internal errors after the rescaling described in Sect.~\ref{sec:internal_errors} are $18$ and $25$ K on \teff, and $0.04$ and $0.05$ dex on \feh for ELODIE and MILES, respectively, which is about $1.25$ times lower than the external dispersions estimated here. Because these dispersions include both the errors on our and on the literature measurements, we conclude that our estimates of the internal errors properly reflect the external errors, or accuracy of the measurements.

\subsection{Results}\label{sec:results}

\begin{figure*}
\begin{center}
	\includegraphics[width=1\textwidth, trim=0.3in 0.in 0.in 0.in]{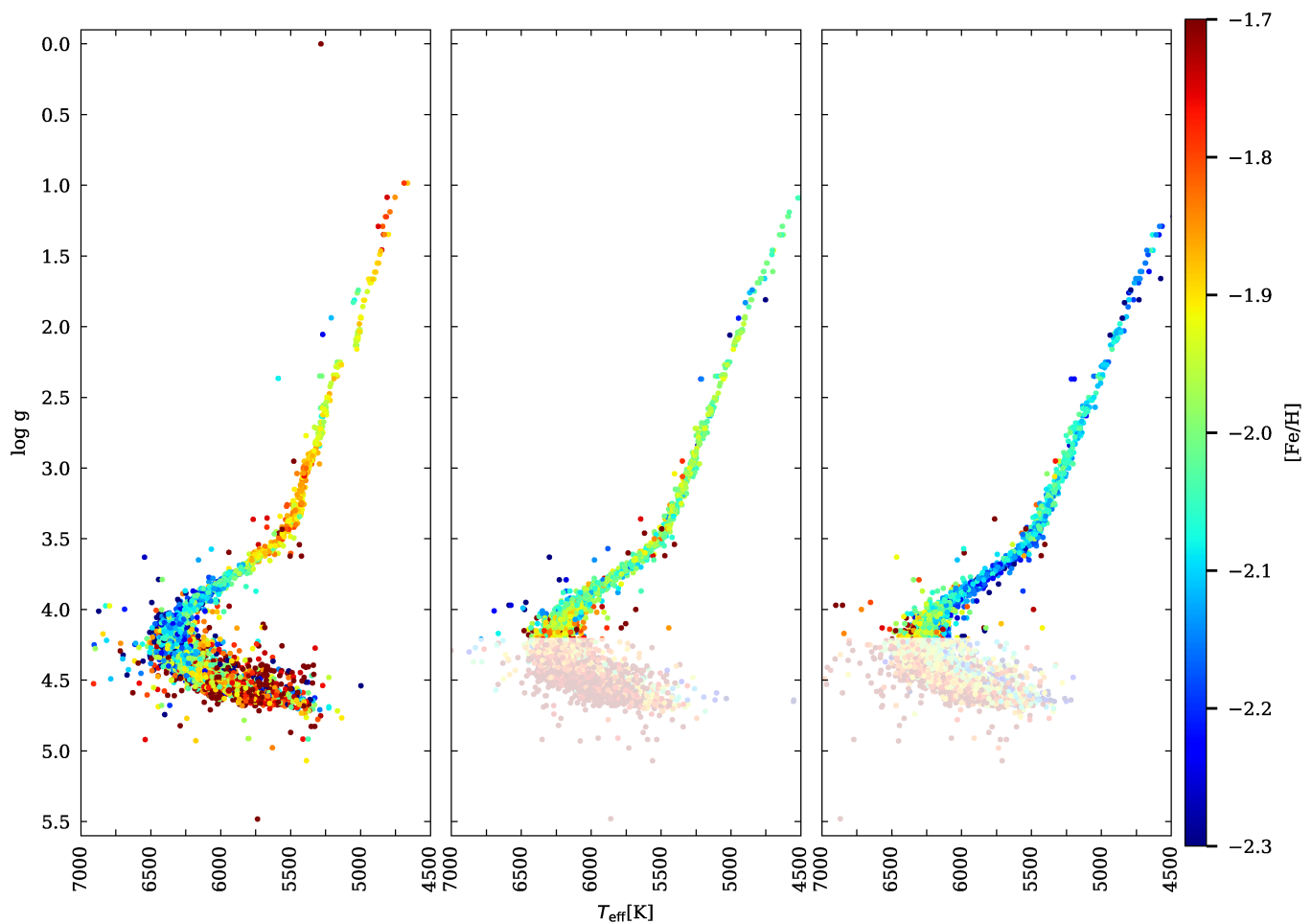}
  \caption{Distribution of the parameter measurements for the 5\,400 spectra of cluster members with good-quality spectra in the \logg vs. \teff plane.
    The left panel shows the \TOH catalogue, the central and right represent our reanalysis with the ELODIE and MILES interpolators, respectively.
    The color scale, shows the metallicity, in the right margin.
    The region of the diagram that we do not discuss here ($\logg < 4.2$) is shown with paler colors.
     }
	\label{fig:hrd}
\end{center}
\end{figure*}

Our measurements of the atmospheric parameters
for the 1\,587 spectra are available in electronic form in Vizier.

Figure \ref{fig:hrd} presents the distribution of the measurements in the \logg versus \teff plane. The vertical axis is the photometric gravity, and the horizontal axis is the \teff from \TOH, ELODIE, and MILES, for the left, central, and right panel, respectively. The data points are color-coded according to the metallicity. This can be compared with the Figure 8 of \TOH.
The RGB harbors an unphysical break in the \TOH data near $\teff = 5000$\,K. We cannot provide a final explanation for this break, but it may correspond to a discontinuity of the opacities used for GSL, at a threshold temperature where some molecular bands are considered. This RGB is generally hotter than in our analysis, reflecting the systematic difference quantified in Sect.~\ref{sec:external} that was also found when \TOH was compared to earlier measurements. The small \teff difference between ELODIE and MILES is consistent with the uncertainties on the temperature calibration of the RGB.

\begin{figure}
\begin{center}
	\includegraphics[width=0.53\textwidth, trim=0.3in 0.in 0.in 0.in]{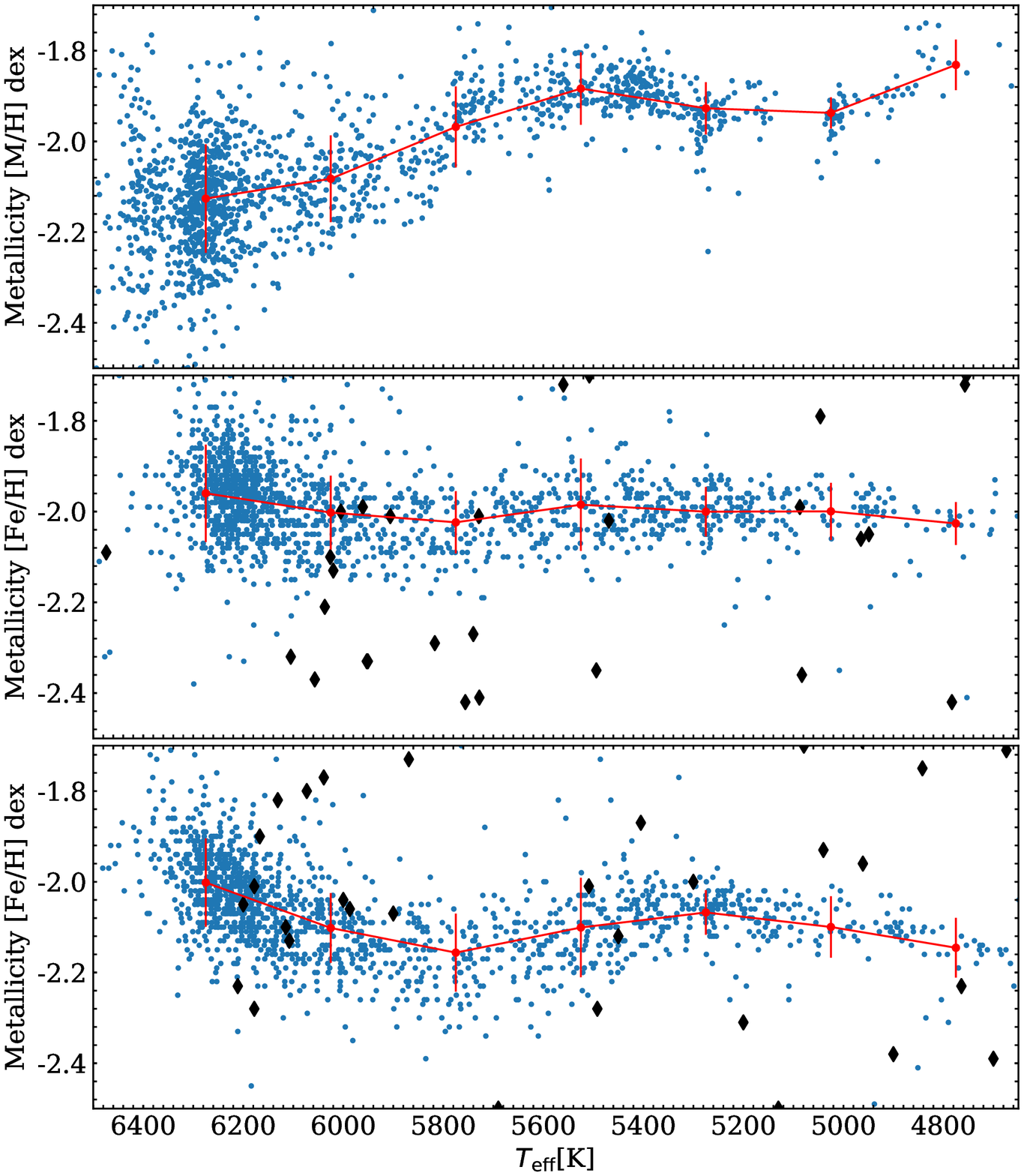}
  \caption{\feh vs. \teff for the giants ($\logg < 4.2$ dex). The top panel shows the \TOH catalog, and the middle and bottom panel represent our reanalysis with the ELODIE and MILES interpolators,respectively (blue dots).
    The solid red line connects the mean values and standard deviations of the metallicity in 250 K wide bins. The black dots show the location of the ELODIE and MILES library stars, in the middle and bottom panel, respectively. The mean [Fe/H] dispersion in \feh of stars taken in bins of 250 K for H16 is 0.076, for ELODIE is 0.075 and that for MILES is 0.079 calculated for stars with \teff$ < 6\,500$K.
         }
	\label{fig:fehteff}
\end{center}
\end{figure}

Figure \ref{fig:fehteff} shows the distribution of the sample in the \feh versus \teff plane. The top panel shows the result from the \TOH analysis, and the middle and lower panels are the results with the ELODIE, and MILES interpolators respectively.
In the three cases, the low dispersion around the observed trend reveals a high internal consistency. The mean \feh dispersion around the cluster isochrone, estimated as the average rms computed in the \teff bins shown in Fig.~\ref{fig:fehteff}, is about $0.07$~dex. Compared to the $0.05$~dex external error reported in Sect.~\ref{sec:external}, we see no  clear indication of a cosmic dispersion.

In addition to  this internal consistency, the three analyses display different patterns of metallicity variation. The results with ELODIE would be consistent with a constant metallicity. The \TOH metallicity steadily increases along the sGB, while the MILES trend is wavy, with a pronounced \feh decline immediately above the TO.

\section{Discussion}\label{sec:discuss}

\subsection{Changes in surface metallicity above the turn-off}

Considering that chemical tagging (the ability to associate individual stars with their birth environment) has been a science driver for many of the large surveys now underway, the ability to differentiate surface from initial abundances is a crucial necessity of stellar evolution models. Many processes can alter the surface abundance of a star during its lifetime. Atomic diffusion is one of the most frequent discussed processes. 
It encompasses different processes that operate in the radiative regions of the stars and cause the redistribution of elements, mainly by the migration of heavier elements toward its center, which consequently lowers the surface metallicity of the stars. Although atomic diffusion in stars has been theoretically predicted more than a century ago by \citet{chapman1917}, its real effect on stellar evolution remains debated \citep{vandenberg2002}.
Mixing processes (e.g., convection and turbulent mixing), compete with gravitational settling to weaken the actual depletion in heavy elements. 
The efficiency of these processes, at variance with the pure atomic diffusion, has no well-established physical description, and can only be determined through comparison with observations. For instance, \citet{dotter2017} investigated models with and without these effects and found that the differences in ages obtained through isochrones might reach up to 20\%. Star clusters, with a coeval stellar population that in principle is initially chemically homogeneous, are the perfect laboratory for testing the balance between the different phenomena.

It is expected that because of atomic diffusion, the surface abundance of heavy element is minimum at the TO, increasing both down the MS and toward the RGB. Metallicity trends  near the TO of globular clusters have previously been discussed in NGC\,6752 by \citet{gruyters2013}, and in M\,30 \citet{gruyters2016}, and in the M\,67 open cluster by \citet{bertelli2018}, \citet{souto2018}, \citet{gao2018}, and \citet{liu2019}. In particular \citet{gratton2001} and more recently  \citet{lovisi2012} found no difference between the abundances of TO stars and base-RGB stars for NGC\,6397.In contrast, \citet{korn2007}, \citet{lind2008} and \citet{nordlander2012} found a clear trend.
The latter three works found similar trends for the iron metallicity from the TO to the RGB, obtaining maximum metallicity differences of about 0.15 $-$ 0.20 dex. They argued that their results indicate the need of atomic diffusion with weak efficiency of turbulent mixing, but also that, based on the uncertainties from modeling techniques, it cannot be ruled out that this might be just an artifact. For example, the Figure 7 of \citet{lind2008} shows that although they and \citet{korn2007} found a very similar trend in [Fe/H] along the Hertzsprung-Ruseel diagram, the systematic differences between their values is about 0.15 dex, which they explained as possible differences between the NLTE corrections. 
In other clusters, the  measured amplitude of variations is smaller, or no effect was detected.
All these works attempted to determine chemical abundances with dedicated techniques, addressing and trying to take into account model weaknesses such as the NLTE and 3D effects. 

We here found different \feh trends above the TO using three different stellar libraries. This casts some doubt on the reality the reality of the detection of atomic diffusion effects.

Regarding the \TOH results, we recall that the magnitude of the trend is significantly smaller than the GSL grid mesh and that the synthetic spectra themselves can introduce systematics due to calculation details. The nonuniform distribution of the cluster stars along the sequence in the the top panel of Fig. 3 is an indication that some finite-grid effects are indeed present along the temperature axis in this analysis, but the ramification for other parameters remain unclear. In contrast, the ELODIE and MILES interpolators do not suffer from this wide-mesh drawback, but they are limited by the small number of stars in this region of the parameter space (about 40 stars  for both libraries in the $\pm 0.3$~dex slice around the cluster metallicity, see Fig.~\ref{fig:fehteff}). These observed library spectra are also subject to noise and to the effect of peculiarities of individual stars. In well-populated regions of the parameter space (typically close to solar metallicity), these effects are averaged out because a fair number of stars have similar parameters, but in this low-metallicity regime, every single star might have a strong influence on the interpolator.

The distribution of the ELODIE stars (the black dots in Fig.~\ref{fig:fehteff}) is uniform except in the region $\teff > 6100$\,K, which includes only one star with $\feh < -1.7$ dex. In the case of MILES, we note that in the range $5500 < \teff < 5850$\,K one library star has $\feh < -1.7$ dex, HD\,140283, with $\feh \approx -2.57$ dex. The MILES pattern in this regime may therefore be an artifact of the interpolation. 
The interpolator we used here is an improvement on the interpolator that was initially published by \citet{prugniel2011}. The improvement concerned the cool stars ($\teff < 4800 K$), and consisted of refining the stellar parameters by critically scrutinizing the literature and tuning the interpolator.
The number of terms in the polynomial development was also increased, essentially to reflect the variation between the giants and the dwarfs better, and this can affect its behavior at warmer temperature. We repeated the analysis with this first version interpolator, in order to verify the reliability of the metallicity trend with respect to the change of the interpolator, and we did not notice any significant difference. During the preparation of this improved MILES interpolator, a series of interpolators were constructed, where for each a single star was excluded. They are called X-interpolators in \citet{sharma16}, and they were used to verify the stability of the results. 
We repeated the analysis here with all the X-interpolators of the library stars in the region of the parameters we analyzed to determine whether the trend was modified for some of these stars. We found the trend to be robust, and therefore rule out any strong bias due to an individual star.
We also compared the parameters that were used to compute the MILES interpolator with an updated compilation of the literature, and did not find any discrepancy. Finally, we redetermined the parameters of the MILES spectra by fitting in the same way as the MUSE observation (self-inversion of the interpolator), and we found no systematic effect there either. 

The difference between the trends found for the three analyses remains to be understood. Each analysis approach has its merits and drawbacks, and we found no reason prefer one more that the others.

\subsection{Spectroscopic surface gravity}\label{sec:freeg}

\begin{figure}
\begin{center}

    \includegraphics[scale=.54]{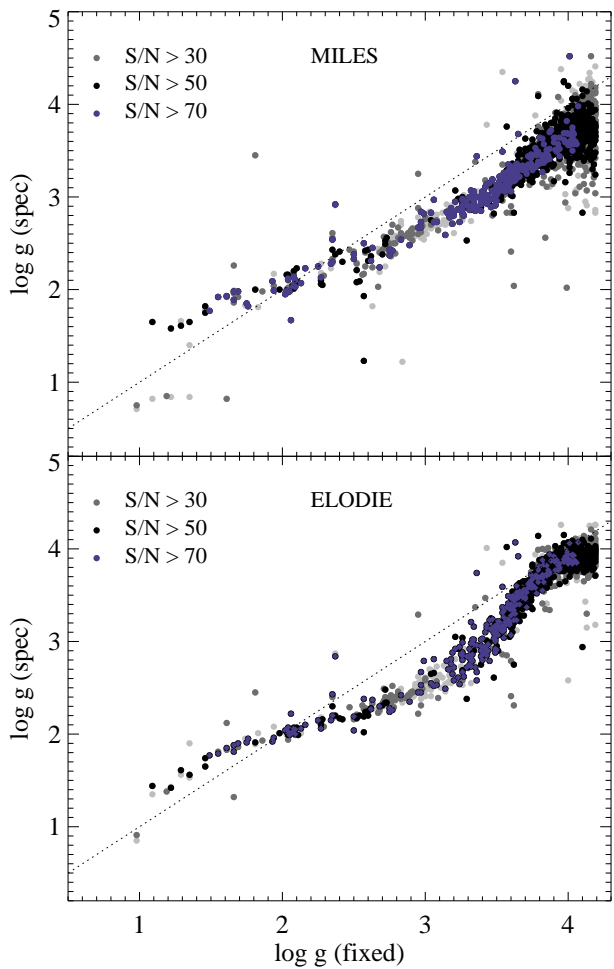} 
        \caption{Comparison between the spectroscopic and photometric \logg. The dotted line is the 1:1 correspondence, the color of the symbol reflects S/R as indicated in the figure. The top panel shows the MILES analysis and the bottom panel that of ELODIE.}
         	\label{fig:logphot_free}
\end{center}
\end{figure}

\begin{figure}
\begin{center}
	\includegraphics[width=0.53\textwidth, trim=0.3in 0.in 0.in 0.in]{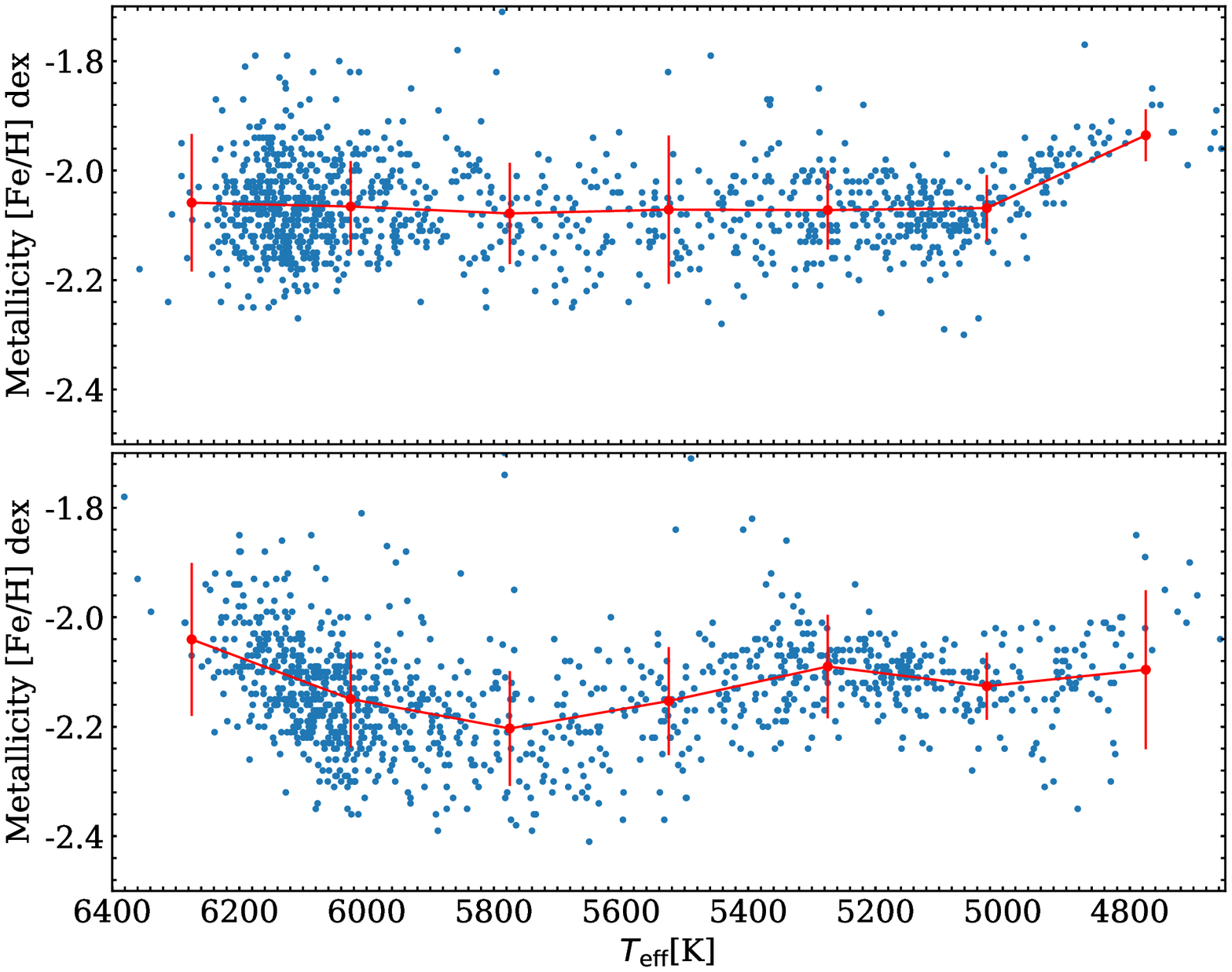}
  \caption{\feh vs. \teff for the giants ($\logg < 4.2$ dex), with \logg determined spectroscopically. The top panel shows results obtained with the ELODIE interpolator, and the bottom panel shows those obtained with the MILES interpolator. The solid red line connects the mean values and standard deviations of the metallicity in 250 K wide bins. The difference with Fig.~\ref{fig:fehteff} is that \logg was this time determined by the full-spectrum fitting, not by photometry.
         }
	\label{fig:fehteff_freeg}
\end{center}
\end{figure}

\TOH did not fit \logg from the spectra because of concerns of degeneracy between the instrumental broadening and the surface gravity. Because the MUSE LSF cannot be precisely assessed for the spectra extracted from the final cubes, the authors preferred to adopt \logg from their photometric fit to the isochrone.
Although the line width is the main indicator for the gravity in individual spectral lines that are observed at high spectral resolution, the degeneracy between the broadening and the gravity may not be important at medium resolution and for full-spectrum fitting, where all the lines are used, notwithstanding their sensitivity to gravity.

To determine the effect of the gravity on the results for the metallicity, we performed a separate analysis with \logg as an additional free parameter.
Figure~\ref{fig:logphot_free} compares the two values. For the high S/N spectra, (S/N$ > 50$), the differences in \logg do not exceed 0.5 dex between the two methods.
When we compare the new values with literature values (see Sect.~~\ref{sec:external}), we find  $\logg({\rm\scriptstyle lit}) - \logg({\rm\scriptstyle spectro}) = 0.15$~dex, with a dispersion of 0.25~dex. \teff obtained with free \logg is lower by $\sim 40$~K (ELODIE) and $85$~K (MILES) compared to the values when \logg was fixed. 

The \feh trends with this analysis, presented in Fig.~\ref{fig:fehteff_freeg}, are not significantly modified in this new analysis (compare with  Fig.~\ref{fig:fehteff}).

\section{Conclusions}\label{sec:conclusions}

We repeated the analysis of 1\,587 MUSE spectra of 1\,063 stars located above the TO of NGC\,6397 that was originally performed by \TOH. While their analysis used a grid of synthetic stellar spectra, we have used two empirical stellar libraries, ELODIE and MILES, to determine \teff and \feh
of these stars. These spectra have $S/N \gtrsim 50$, and comparisons with \TOH and earlier studies suggest that the errors 
in our determinations are $\sim 25$ K and $0.05$ dex on the two parameters respectively. The metallicity dispersions of the measurements, $\sim 0.07$ dex, agree with these figures.

Despite this fair consistency, the metallicity trends along the sGB and RGB are different in the three analyses. 
No trend was found with ELODIE, and the trend found with MILES does not match the theoretical expectations.
Previous observational studies of this and other clusters also lead to contradictory results about the reality of this variation, and
the detection of the effect of atomic diffusion.

These empirical libraries contain only a limited number of stars in this low-metallicity regime (they do not have enough stars on the MS, which makes it impossible to explore the trend below the TO).
Despite this limitation, the empirical libraries are essential for validating the synthetic libraries, and they are still required ingredients for models of integrated spectra of stellar populations as are used in extragalactic astronomy. Therefore, the motivation to continue improving the empirical libraries remains strong. We here clearly identified the regime of \feh $\sim -2$ as an area where the libraries deserve to be improved.

The metallicity trend found by \TOH qualitatively reproduces the expected effect of atomic diffusion. The reference grid of synthetic spectra has a wide metallicity mesh, however, which casts doubts on the capability of resolving the small physical effect. 
When the grid mesh are compared to the generally accepted accuracy in the determinated atmospheric parameters, most of the grids that are currently used for population analysis are too coarse in their metallicity coverage.
For $\teff \sim 6\,000$~K, the mesh size typically is 200 K, which is roughly four times the admitted precision on measurements. On \logg the size is 0.5~dex, which about five times the precision, and on \feh, the size is 0.5~dex, which is about ten times the precision. For example, for the GSL the meshes sizes are 100~K, 0.5, and 0.5~dex, and in \citet{coelho2014} they were 200~K, 0.5, and 0.5~dex on the three parameters. Only recently have libraries with a finer metallicity sampling become available. For example, INTRIGOS \citep{franchini2018}  and \citet{allendeprieto2018} have mesh sizes of 250~K, 0.5, and 0.25~dex. Adopting finer sampling is naturally challenging in terms of both the computing time required to produce the large number of spectra, and the data volume which needs to be manipulated, which is still in the 100s GB in \citet{allendeprieto2018}.
To properly resolve the metallicity trends due to atomic diffusion, or to determine the other surface abundances alterations due to other processes, a mesh size of 0.1~dex is required.

From point of view of high resolution abundance analysis, there is also large space for improvement. Limitations such as the treatment of NLTE and 3D effects are still a large source of uncertainty. 

We illustrated the limits of the current libraries at low metallicity. While observations of Galactic globular clusters are classical benchmark for population models, the empirical libraries are at best limited near the TO of these clusters, and the interpolated spectra used for the models are crudely approximate. The most recent library is the X-Shooter Spectral Library, XSL \cite{arentsen2019}. It has a more uniform coverage of the metallicity range than earlier libraries, but also suffers from a relative shortage of TO and MS stars in the low metallicity regime. For the stellar population models and for the validation of synthetic spectra, it is desirable to extend the coverage of these libraries.

\begin{acknowledgements}
We thank Tim-Oliver Husser who shared with us the tables with the results of the H16 analysis, and the MUSE team for making public the extracted spectra on their web site.
We acknowlege the LABEX Lyon Institute of Origins (ANR-10-LABX-0066) of the Université de Lyon for its financial support within the program "Investissements d'Avenir" (ANR-11-IDEX-0007) of the French government operated by the National Research Agency (ANR), and IDEXLYON for their support fo the Indo-French Astronomy school (IFAS4), where this project was started. This work was supported by the Programme National Cosmology et Galaxies (PNCG) of CNRS$/$INSU with INP and IN2P3, co-funded by CEA and CNES.
We thank Yasna Ordenes, Yue Wu, Jean Damascène Mbarubucyeye, Sina Chen, Sonali Borah, Raghu Prasad, and Saumya Gupta, who contributed to early investigations on this project.
We thank the referee for the comments, which helped improve the quality of the paper. 
L.P.M. thanks FAPESP (grant 2018/26381-4) and CNPQ (grant 306359/2018-9). L.P.M. also thanks L.Tresse, director of CRAL, for the hosting during the development of this work. 

\end{acknowledgements}
\bibliographystyle{aa} 
\bibliography{paper} 
%
%
%
%
%
%
%
\end{document}